\documentclass[aps,prb,amsfonts,amssymb,twocolumn,amsmath,preprintnumbers,nofootinbib,floatfix,superscriptaddress,
showpacs]{revtex4-1}
\usepackage[dvips]{graphics}
\usepackage{graphicx}
\usepackage{bm}

\begin{document}

\title{Enhancing of the in-plane FFLO-state critical temperature in heterostructures by the orbital effect of the magnetic field}
\author{A. M. Bobkov}
\affiliation{Institute of Solid State Physics RAS, Chernogolovka,
Moscow reg., 142432 Russia}
\author{I. V. Bobkova}
\affiliation{Institute of Solid State Physics RAS, Chernogolovka,
Moscow reg., 142432 Russia}
\affiliation{Moscow Institute of Physics and Technology, Dolgoprudny, 141700 Russia}

\date{\today}

\begin{abstract}
It is well-known that the orbital effect of the magnetic field suppresses superconducting $T_c$. We show that for a system, which is in the Larkin-Ovchinnikov-Fulde-Ferrell (FFLO) state at zero external magnetic field, the orbital effect of an applied magnetic field can lead to the enhancement of the critical temperature higher than $T_c$ at zero field. We concentrate on two systems, where the in-plane FFLO-state was predicted recently. These are equilibrium S/F bilayers and S/N bilayers under nonequilibrium quasiparticle distribution. However, it is suggested that such an effect can take place for any plane superconducting heterostructure, which is in the in-plane FFLO-state (or is close enough to it) at zero applied field.
\end{abstract}

\maketitle

\section{introduction}

Usually magnetic fields suppress superconductivity. The enhancement or, at least, recovery of superconductivity by a magnetic field is a counterintuitive phenomenon. There are two mechanisms of superconductivity
destruction by a magnetic field: orbital effect and Zeeman interaction of electron spins with
the magnetic field. In the present paper we consider the influence of the orbital effect of the magnetic field on superconductivity. The orbital (or electromagnetic) effect of the magnetic field causes the Meissner currents in superconductors. It is widely accepted that the orbital effect is always destructive to superconductivity. 

It is worth noting here that quantum effects of an electron motion in a magnetic field can
result in the appearance of so-called "reentrant superconductivity" with $dT_c/dH>0$ in layered and isotropic three-dimensional
type-II superconductors\cite{lebed86}-\cite{rasolt87}. However, even in these cases the maximal $T_c$ cannot exceed the critical temperature of the system at zero applied field. 

Here we demonstrate that the superconducting critical temperature ($T_c$) can be increased by the orbital effect of the applied magnetic field. The point is that we can have enhancement of superconductivity if, due to any reasons, we have the first-order correction (in the magnetic field) to the condensate wave function. It is well known that in the standard case there is no first-order correction to the condensate wave function because the only possible scalar quantity is $div \bm A=0$. The first-order correction may exist if there is a distinguished vector in the system without external field. In this case one can have another scalar quantity containing the magnetic field in the first approximation. We have found an example - the superconducting heterostructure in the Fulde-Ferrell-Larkin-Ovchinnikov (FFLO) state - where this possibility is realized. More particular, we show that in a S/F or S/N heterostructure, which is in the FFLO state at zero external field, the orbital effect of a weak applied magnetic field can lead to the enhancement of the critical temperature higher than $T_c$ at zero field. 

In the FFLO-state \cite{larkin64,fulde64} the pairing is realized by electrons
with opposite spins, which do not have zero total
momentum anymore due to the Zeeman splitting. Therefore, the Cooper pair as a whole acquires total momentum $\bm k$. For the considered here heterostructures the role of the distinguished vector, discussed above, is played by the quantity $\bm n \times \bm k$, where $\bm n$ is the unit vector normal to the bilayer plane.    

Experimentally the FFLO-state is usually expected to be observed for superconductors in an applied magnetic field if the  Zeeman interactions of electron spins with a magnetic field dominates the orbital pair breaking (systems with large
effective mass of electrons \cite{bianchi03,capan04}, thin films and layered superconductors
under in-plane magnetic field \cite{uji01}). Now there is a growing body of experimental evidence for the FFLO phase, generated by the applied magnetic field, reported from various measurements \cite{singleton00}-\cite{uji12}. 

The effect discussed in the present paper does not occur if the FFLO-state is generated by an applied magnetic field. However, there are several types of systems, where the FFLO-state can be realized without an external magnetic field. Here we concentrate on two possible systems, where the FFLO-state was predicted recently. These are S/F heterostructures \cite{mironov12} and S/N heterostructures under nonequilibrium quasiparticle distribution \cite{bobkova13}. It should be noted that in these heterostructures the FFLO state can occur in the dirty limit and, more important, for temperatures close to $T_c$, though usually it is fragile with respect to disorder and can exist only at low enough temperatures. It is also worth to underline that this is the so-called in-plane FFLO-state, where the superconducting order parameter profile is modulated along the layers. It should be distinguished from
the normal to the S/F interface oscillations of the condensate wave function in the ferromagnetic layer, which are well investigated as theoretically, so as experimentally \cite{buzdin05,bergeret05,sidorenko09}. It is worth to note that there are other proposals of the FFLO-state without applied magnetic field, where the discussed here effect could be also related. Among them a current-driven FFLO-state in 2D superconductors with Fermi surface nesting \cite{doh06} and a FFLO-state in unconventional superconducting films \cite{vorontsov09}.

\section{model and calculation procedure}

Now we proceed with the microscopic calculations of $T_c$ of the S/F bilayer in the FFLO-state under the applied magnetic field. As it was shown in \cite{mironov12}, under the appropriate parameters the FFLO-state can be realized in this system close to $T_c$ at zero applied field. The particular parameters of the bilayer are given in the caption to Fig.~\ref{TcH1}. Let us assume that a weak external magnetic field $\bm H$ is applied in the plane of the bilayer [($yz$)-plane]. We also choose the vector potential $\bm A=(0, H_z x, -H_y x)+\bm A_0$ to be parallel to the ($yz$)-plane. In our calculations we assume that (i) S is a singlet s-wave superconductor; (ii) the system is in the dirty limit, so the quasiclassical Green's function obeys Usadel equations \cite{usadel}; (iii) the thickness of the S layer $d_S \lesssim \xi_S$. Here $\xi_S=\sqrt{D_S/\Delta_0}$ is the superconducting coherence length, $D_S$ is the diffusion constant in the superconductor and $\Delta_0$ is the bulk value of the superconducting order parameter at zero temperature. This condition allows us to neglect the variations of the superconducting order parameter and the Green's functions across the S layer; (iv) we work in the vicinity of the critical temperature, so the Usadel equations can be linearized with respect to the anomalous Green's function.

The retarded anomalous Green's function $\hat f^R \equiv \hat f^R(\varepsilon,\bm r)$ depends on the quasiparticle energy $\varepsilon$ and the coordinate vector $\bm r=(x, \bm r_\parallel)$, where $x$ is the coordinate normal to the S/F interface and $\bm r_\parallel$ is parallel to the interface [$(yz)$-plane]. $\hat ~$ means that the anomalous Green's function is a $2\times 2$ matrix in spin space. We assume that the exchange field in the F layer is homogeneous $\bm h=(0,0,h)$ \cite{note_zeeman}. In this case there are only singlet and triplet with zero spin projection on the quantization axis pairs in the system. In the language of Pauli matrices it means that $\hat f^R(\varepsilon,\bm r)=[f_\uparrow (1+\sigma_3)/2+f_\downarrow (1-\sigma_3)/2]i\sigma_2$, where $\sigma_{2,3}$ are the corresponding Pauli matrices in spin space. While we only consider the singlet pairing channel, the superconducting order parameter $\hat \Delta=\Delta i \sigma_2$.

The linearized Usadel equation for the retarded anomalous Green's function $f^R_\sigma$, where $\sigma=\uparrow,\downarrow$, takes the form:
\begin{equation}
D (\nabla-\frac{2ie}{c}\bm A)^2 \hat f^R_\sigma + 2 i (\varepsilon+\sigma h) \hat f^R_\sigma +2 \pi \hat \Delta = 0
\enspace .
\label{usadel}
\end{equation}
Here $\sigma h=+h(-h)$ for $f_{\uparrow(\downarrow)}$. $D$ stands for the diffusion constant, which is equal to $D_{S(F)}$ in the superconductor (ferromagnet).

Eq.~(\ref{usadel}) should be supplied by the Kupriyanov-Lukichev boundary conditions \cite{kupriyanov88} at the S/F interface ($x=0$)
\begin{eqnarray}
\sigma_S \partial_x f^R_{\sigma,S} = \sigma_F \partial_x f^R_{\sigma,F} = g_{FS}\left.(f^R_{\sigma,S}-f^R_{\sigma,F})\right|_{x=0}
\label{interface_cond}
\enspace ,
\end{eqnarray}
where $\sigma_{S(F)}$ stands for a conductivity of the S(F) layer and $g_{FS}$ is the conductance of the S/F interface. The boundary conditions at the ends of the bilayer are $\left. \partial_x f^R_{\sigma,S} \right|_{x=d_S} = \left. \partial_x f^R_{\sigma,F} \right|_{x=-d_F}=0 $.

In the FFLO-state the superconducting order parameter and the anomalous Green's function are spatially modulated. We assume that
$\Delta(\bm r)=\Delta\exp(i\bm k\bm r_\parallel)$ and $f(\bm r)=f(x)\exp(i\bm k \bm r_\parallel)$. Substituting the modulated Green's function into the Usadel equation and solving it under the assumption, that the anomalous Green's function weakly varies across the S layer, we obtain the anomalous Green's functions in the bilayer. We calculate the anomalous Green's function up to the second order in the magnetic field. In the S layer the corresponding zero and first order terms take the form:
\begin{equation}
f_{\sigma,S}^{(0)}=\frac{i\pi\Delta}{E}
\enspace ,
\label{fS0}
\end{equation}
\begin{eqnarray}
f_{\sigma,S}^{(1)}=\frac{2ieD_S}{cd_SE}\left[ f^{(0)}_{\sigma,S} \int \limits_0^{d_S} dx \bm k \bm A(x) + \right.~~~~~~ \nonumber \\
\left. \frac{g_{FS}\int \limits_{-d_F}^0 dx f^{(0)}_{\sigma,F}(x)\bm k \bm A(x) \frac{\cosh[\lambda_\sigma(x+d_F)]}{\cosh[\lambda_\sigma d_F]}}{\sigma_S(\lambda_\sigma \tanh[\lambda_\sigma d_F]+g_{FS}/\sigma_F)} \right], \label{fS1}
\end{eqnarray}
\begin{equation} 
E=\varepsilon+iD_Sk^2/2 + \frac{i g_{FS}D_S \lambda_\sigma \tanh[\lambda_\sigma d_F]}{2\sigma_S d_S (\lambda_\sigma \tanh[\lambda_\sigma  d_F]+g_{FS}/\sigma_F)} \nonumber
\enspace .
\end{equation}
In Eq.~(\ref{fS1}) $\lambda_\sigma^2=k^2-2i(\varepsilon+\sigma h)/D_F $ and $f^{(0)}_{\sigma,F}(x)$ is the zero-field anomalous Green's function in the F layer, which takes the form
\begin{equation}
f_{\sigma,F}^{(0)}(x)=\frac{(g_{FS}/\sigma_F)\cosh[\lambda_\sigma(x+d_F)]}{\lambda_\sigma \sinh [\lambda_\sigma d_F]+(g_{FS}/\sigma_F)\cosh[\lambda_\sigma d_F]}f_{\sigma,S}^{(0)}
\enspace .
\label{fN0}
\end{equation}

The second-order term $f_{\sigma,S}^{(2)}$ of the anomalous Green's function in the superconducting layer of the F/S bilayer takes the form:

\begin{equation}
f_{\sigma,S}^{(2)}=\frac{2iD_S}{d_SE}\left[ M_{\sigma,S} + \frac{g_{FS}\int \limits_{-d_F}^0 dx M_\sigma (x) \frac{\cosh[\lambda_\sigma(x+d_F)]}{\cosh[\lambda_\sigma d_F]}}{\sigma_S(\lambda_\sigma \tanh[\lambda_\sigma d_F]+g_{FS}/\sigma_F)} \right] 
\enspace ,
\label{fs2}
\end{equation}
where $ E $ is defined below Eq.~(4). $M_{\sigma,S}$ and $M_\sigma(x)$ are the following expressions
\begin{equation}
M_{\sigma,S}=\frac{e}{c}f_{\sigma,S}^{(1)}\int \limits_0^{d_s}dx \bm k \bm A(x)-\frac{e^2}{c^2}f_{\sigma,S}^{(0)}\int \limits_0^{d_s}dx  \bm A^2(x)
\enspace ,
\end{equation} 
\begin{equation}
M_{\sigma}(x)=\frac{e}{c}\bm k \bm A(x) f_{\sigma,F}^{(1)}(x)-\frac{e^2}{c^2}\bm A^2(x)f_{\sigma,F}^{(0)}(x)
\enspace ,
\end{equation}
where the first-order term $f_{\sigma,F}^{(1)}(x)$ in the ferromagnet is required. It takes the form:
\begin{eqnarray}
f_{\sigma,F}^{(1)}(x)=C_\sigma \cosh \left[ \lambda_\sigma (x+d_F) \right]- \nonumber \\
\frac{4e}{c}\int \limits_{-d_F}^x dx' \sinh \left[ \lambda_\sigma (x-x') \right] \bm k \bm A(x') \frac{f_{\sigma,F}^{(0)}(x')}{\lambda_\sigma} 
\enspace ,
\end{eqnarray} 
\begin{eqnarray}
C_\sigma = \frac{\frac{g_{FS}}{\sigma_F}f_{\sigma,S}^{(1)}+\frac{4e}{c}\int \limits_{-d_F}^0 d x L(x)\bm k \bm A(x) f_{\sigma,F}^{(0)}(x)}{\lambda_\sigma \sinh[\lambda_\sigma d_F]+\frac{g_{FS}}{\sigma_F}\cosh[\lambda_\sigma d_F]}
\enspace ,\\
L(x)=\cosh[\lambda_\sigma x]-\frac{g_{FS}}{\sigma_F \lambda_\sigma}\sinh[\lambda_\sigma x] \nonumber
\enspace .
\end{eqnarray}    

In the limiting case of fully transparent S/F interface $g_{FS} \to \infty$ our solution for zero applied field Eq.~(\ref{fS0}) coincides with the one obtained in \cite{mironov12}.

The critical temperature of the bilayer should be determined from the self-consistency equation
\begin{equation}
\Delta=\int \limits_{-\omega_D}^{\omega_D} \frac{d\varepsilon}{4\pi} \Lambda \sum \limits_\sigma {\rm Im} \left[ f_{\sigma,S} \right] \tanh \frac{\varepsilon}{2T}
\label{Tc}
\end{equation}
where $\omega_D$ is the cutoff energy and $\Lambda$ is the dimensionless coupling constant.

\section{results}

It is seen from Eq.~(\ref{fS1}) that if $\bm A \perp \bm k$ (that is $\bm H \parallel \bm k$) the magnetic field does not influence $T_c$ in the linear order. Reversal of the magnetic field is equivalent to reversal of $k$. The enhancement of $T_c$ is maximal if $\bm k$ is antiparallel to $\bm A$ and is replaced by the suppression if $\bm k$ and $\bm A$ are parallel. Now it is worth to note the principal difference between $\bm A$ and $\bm k$. While $\bm A$ is an external controllable parameter, $\bm k$ is an internal parameter of the system. Therefore, two different situations are possible: (i) under applied field the system can adjust the direction of $\bm k$ in order to reach the most energetically favorable configuration. Then $\bm k$ tends to be antiparallel to the vector potential for an arbitrary direction of $\bm A$; (ii) the direction of $\bm k$ is fixed by boundary conditions or other reasons. Then the effect depends on the orientation of the magnetic field with respect to $\bm k$ and is maximal for their antiparallel direction. For this reason further we assume that $\bm k$ and $\bm A$ are aligned along the same direction.

\begin{figure}[!tbh]
   \begin{minipage}[b]{0.5\linewidth}
     \centerline{\includegraphics[clip=true,width=1.69in]{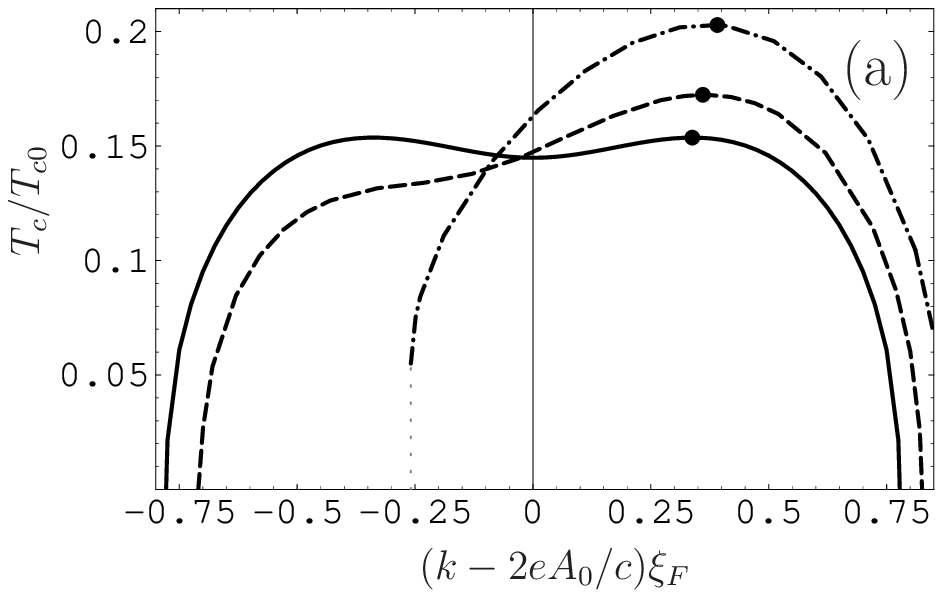}}
     \end{minipage}\hfill
    \begin{minipage}[b]{0.5\linewidth}
   \centerline{\includegraphics[clip=true,width=1.68in]{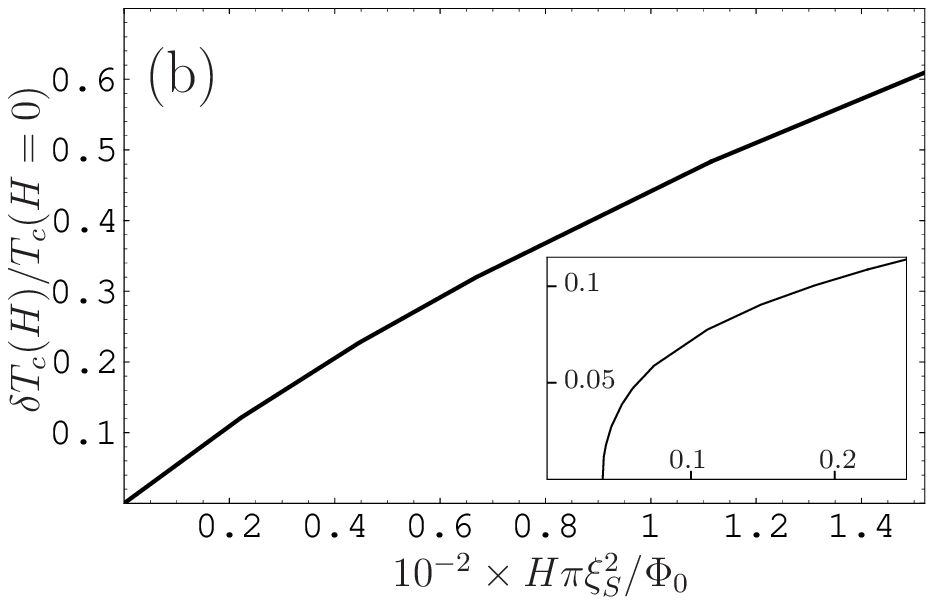}}
  \end{minipage}
   \caption{(a) $T_c$ as a function of $(k-2eA_0/c)$ for the S/F bilayer in the FFLO-state. $T_c$ is measured in units of the superconducting bulk critical temperature $T_{c0}$ and $k$ is measured in units of $\xi_F^{-1}=\sqrt{h/D_F}$. $H=0$ (solid curve); $H=0.0022$ (dashed); $H=0.0067$ (dashed-dotted). (b) $\delta T_c (H)/T_c(H=0)$ as a function of the magnetic field.  The field is measured in units of $\hbar c/e\xi_S^2$. The other parameters of the system are the following: $d_F=1.2\xi_F$, $\sigma_S/\sigma_F=0.5$, $g_{FS}\xi_F/\sigma_F=10$,  $D_S/D_F=0.013$,  $T_{c0}/h=0.1$, $d_S=0.778\xi_S$. Insert: $T_c (H)/T_{c0}$ for superconductivity induced by the field. $d_S=0.769\xi_S$, the other parameters are the same as for panels (a) and (b).}
\label{TcH1}
\end{figure}

The critical temperature of the S/F bilayer as a function of $(k-2eA_0/c)$, calculated according to Eq.~(\ref{Tc}), is represented in Fig.~\ref{TcH1}(a). For the considered problem it is the quantity $(k-2eA_0/c)$ that is gauge invariant, so we represent the curves as functions of this quantity. Different curves correspond to different values of the applied magnetic field $H$. The optimal values of the modulation vector $(k-2eA_0/c)_{opt}$, corresponding to the maximal $T_c$, are marked by points. It is seen that the bilayer is in the FFLO-state in the vicinity of $T_c$ at zero field. The curves corresponding to nonzero fields demonstrate that $T_c(H \neq 0)$ is higher than $T_c(H=0)$. Dependence of $\delta T_c \equiv T_c(H)-T_c(H=0)$ on $H$ is represented in Fig.~\ref{TcH1}(b). It is seen that, if the system is in the FFLO-state at zero field, $T_c$ grows with the applied field  near linearly for weak fields. For the system under consideration the $T_c$ growth is approximately $60\%$ in the given magnetic field range. We do not consider higher magnetic fields because our approximation up to the second order of the field fails to work and one needs to study the exact solution. This is the prospect for future work.

It is worth to note here that the plane wave is not the only possible type of the spatially
modulated FFLO-state at zero field. There can be also stationary wave states modulated as
$\cos \bm k \bm r_\parallel$ and also 2D modulated structures. Simple analysis shows that in the first approximation  $T_c$ of the stationary wave state $\propto \cos \bm k \bm r_\parallel$ is not sensitive to the field. So, it is reasonable to assume that under the applied field the plane wave state should be more energetically favorable than the stationary wave one, at least close to $T_c$. Therefore, we only discuss the plane wave state here.

It is also interesting to note that for a cylinder of a particular radius, made of material in the FFLO-state, one can either increase or decrease $T_c$ by applying the magnetic flux \cite{zyuzin08,zyuzin09}. But this effect is, in fact, a geometrical resonance and also does not lead to the enhancement of $T_c$ higher than $T_c$ for infinite-radius system at zero field.

As it is shown above, $T_c$ of the S/F bilayer in the FFLO-state can be enhanced by the orbital effect of the magnetic field. Let us slightly decrease the width of the S layer in order to completely suppress superconductivity in the bilayer by the exchange field of the F layer, so that the bilayer is in the normal state at $T=0$ and $H=0$. Then superconductivity can be induced by the orbital effect of an external field. This is demonstrated in the insert to Fig.~\ref{TcH1}(b), where $T_c(H=0)=0$.  

It is worth noting here that we do not take into account the Zeeman effect of the applied field. While it is quite reasonable for weak magnetic fields we consider, the Zeeman effect can become important for higher fields. It is well-known that it suppresses  superconductivity. However, this suppression can be compensated by creation of the appropriate nonequilibrium distribution in the bilayer, as it was shown in \cite{bobkova11}. So, in principle, it is possible to observe the discussed effect of $T_c$ enhancement by orbital effect even at large enough fields.

\begin{figure}[!tbh]
   \begin{minipage}[b]{0.5\linewidth}
     \centerline{\includegraphics[clip=true,width=1.69in]{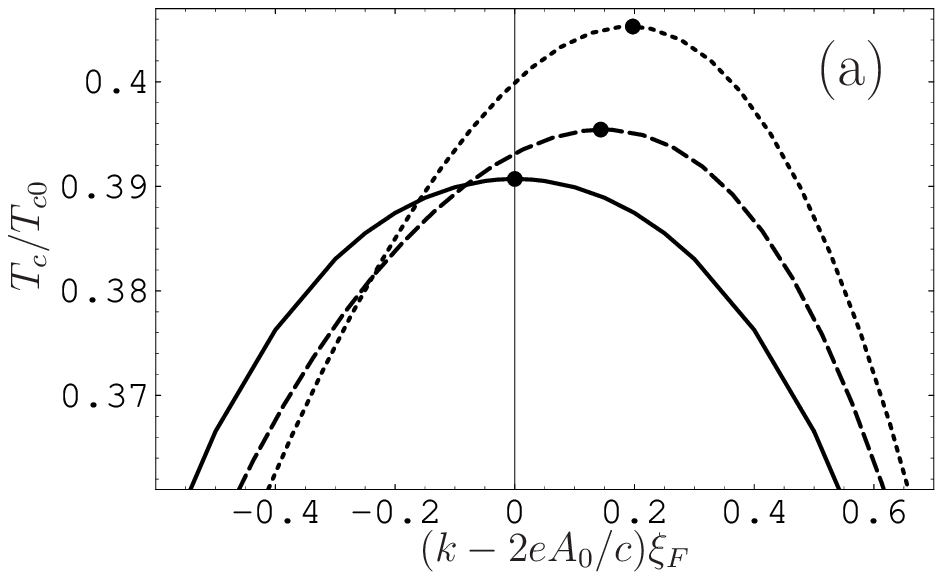}}
     \end{minipage}\hfill
    \begin{minipage}[b]{0.5\linewidth}
   \centerline{\includegraphics[clip=true,width=1.68in]{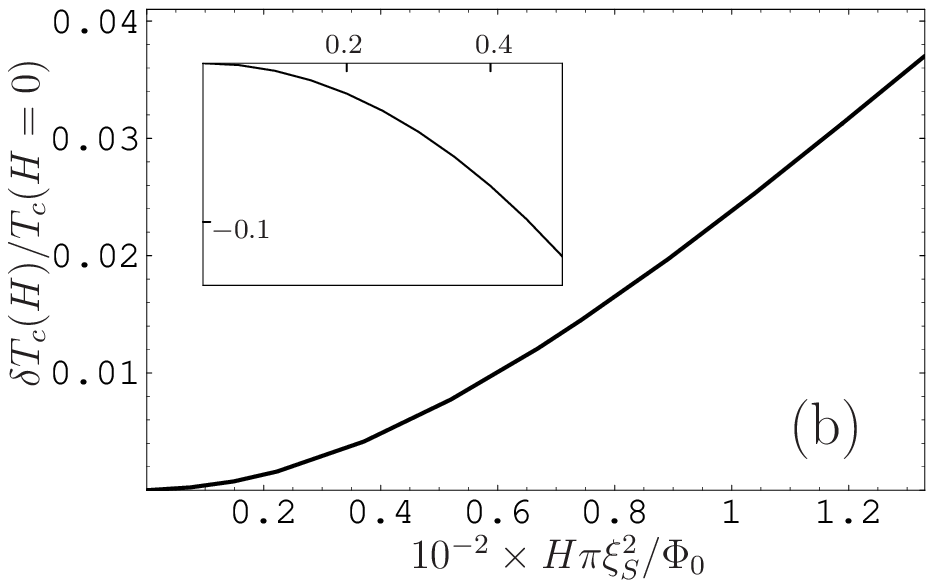}}
  \end{minipage}
   \caption{(a) $T_c$ as a function of $(k-2eA_0/c)$ for the S/F bilayer, which is close to the FFLO-state. $T_c$, $k$ and $H$ are measured in the same units as in Fig.~\ref{TcH1}. $H=0$ (solid curve); $H=0.0067$ (dashed); $H=0.0134$ (dotted). (b) $\delta T_c (H)/T_c(H=0)$ as a function of the magnetic field. $d_S=0.880\xi_S$, the other parameters are the same as in Fig.~\ref{TcH1}. Insert: $\delta T_c(H)/T_c(H=0)$ for the superconducting film.}
\label{TcH2}
\end{figure} 

$T_c$ of the S/F bilayer can also be enhanced by magnetic field even if it is not in the FFLO-state, but it should be rather close to the FFLO-state at zero field. It means that in the bilayer $dT_c/dk^2$ at $k \to 0$ is negative, but it is very small. This situation is demonstrated in Fig.~\ref{TcH2}. $T_c$ as a function of $k$ is represented in Fig.~\ref{TcH2}(a). The bilayer is in the homogeneous state at zero field. In this case $\delta T_c$ rises quadratically
upon increase of the field, as it is plotted in Fig.~\ref{TcH2}(b). While there is no first-order term (in magnetic field) in the anomalous Green's function in this case, the second order term [Eq.~(\ref{fs2})] contains two different contributions, one of which is responsible for the usual suppression of superconductivity and the other one leads to its enhancement. The last contribution overcomes the first one if the system is close enough to the FFLO-state. So, one can conclude that the orbital effect of the magnetic field can be qualitatively different for the S/F bilayer and for the conventional superconducting film, which is not close to the FFLO-state. For comparison in the insert to Fig.~\ref{TcH2}(b) we show $\delta T_c(H)$ for the superconducting film: it behaves in a standard way, that is decreases quadratically upon increase of the field.

\begin{figure}[!tbh]
   \begin{minipage}[b]{0.5\linewidth}
     \centerline{\includegraphics[clip=true,width=1.69in]{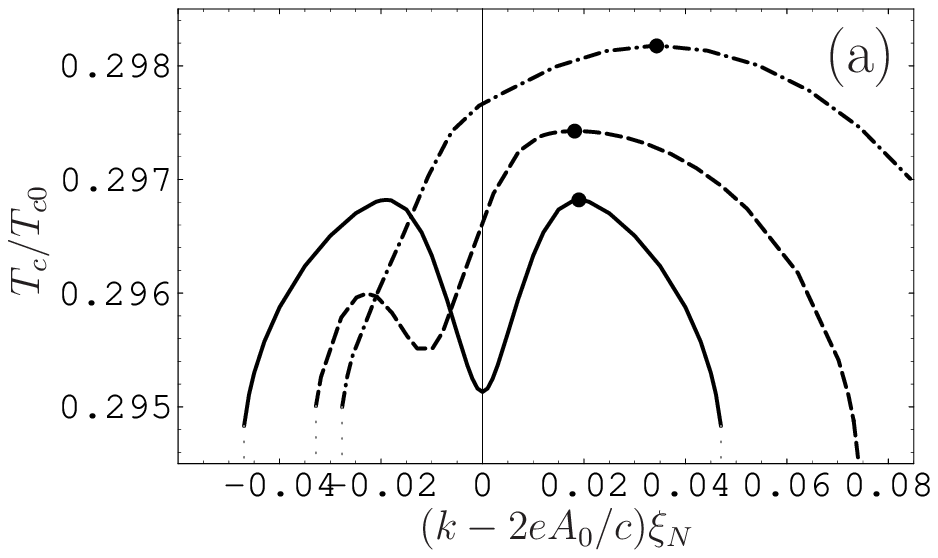}}
     \end{minipage}\hfill
    \begin{minipage}[b]{0.5\linewidth}
   \centerline{\includegraphics[clip=true,width=1.67in]{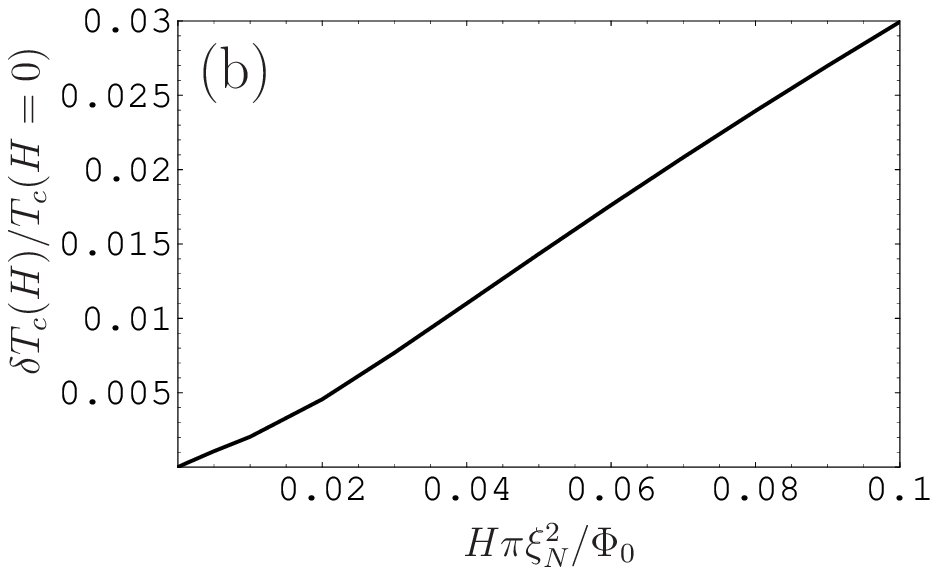}}
  \end{minipage}
   \caption{(a) $T_c$ as a function of $(k-2eA_0/c)$ for the S/N bilayer under nonequilibrium conditions. $T_c$ is measured in units of  $T_{c0}$ and $k$ is measured in units of $\xi_N^{-1}=\sqrt{T_{c0}/D_N}$. $H=0$ (solid curve); $H=0.01$ (dashed); $H=0.02$ (dashed-dotted). (b) $\delta T_c (H)/T_c(H=0)$ as a function of the magnetic field. The field is measured in units of $\hbar c/e\xi_N^2$. The other parameters are the following: $d_N=0.9042\xi_N$, $d_S=0.1205 \xi_N$, $g_{NS}\xi_N/\sigma_S=6.6343$, $\sigma_S/\sigma_N=0.5$, $D_S/D_N=0.04$, $eV/T_{c0}=0.70453$.}
\label{TcH3}
\end{figure}

From the analysis of Eq.~(\ref{fS1}) one can suggest that the orbital effect of the field should lead to $T_c$ enhancement for any system, which is in the FFLO-state (near $T_c$) at zero field. It is because of the general term $\propto \bm k \bm A$ \cite{note_kA}. In order to illustrate this statement we consider another system, where in-plane FFLO-state is possible at zero external magnetic field. This is the S/N bilayer under nonequilibrium quasiparticle distribution \cite{bobkova13}. For this system $T_c$ as a function of $k$ is represented in Fig.~\ref{TcH3}(a). $\delta T_c(H)$ is plotted in Fig.~\ref{TcH3}(b). Qualitatively the situation here is analogous to the S/F bilayer: $\delta T_c(H)$ rises near linearly upon increase of the field. However, this growth is slower than for the S/F bilayer. The reason is that $k_{opt}(H=0)$ is much less in this case. And, in general, the first order term in the anomalous Green's function is $\propto \bm k \bm A$, so the smaller the value of $k_{opt}$, the slower the rate of $T_c$ growth.

\section{discussion}

\begin{figure}[!tbh]
   \begin{minipage}[b]{0.5\linewidth}
     \centerline{\includegraphics[clip=true,width=1.69in]{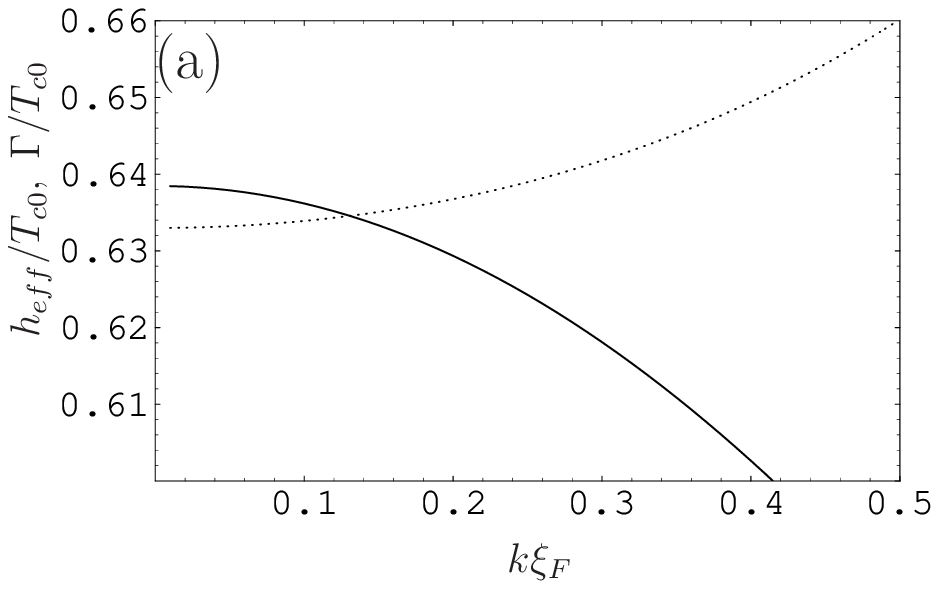}}
     \end{minipage}\hfill
    \begin{minipage}[b]{0.5\linewidth}
   \centerline{\includegraphics[clip=true,width=1.67in]{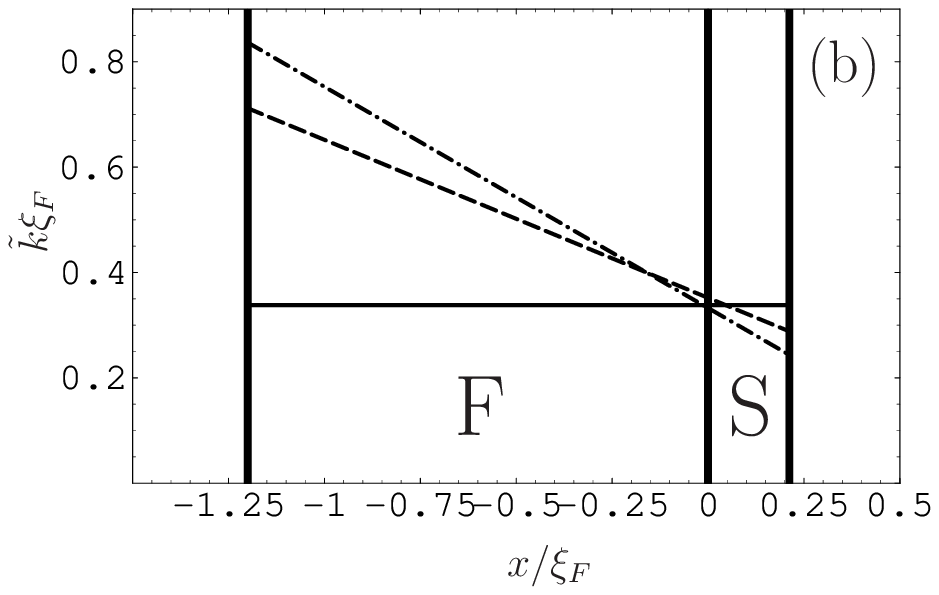}}
  \end{minipage}
   \caption{(a) $h_{eff}$ (solid line) and $\Gamma$ (dotted line) as functions of the pair momentum $k$. (b) Generalized momentum as a function of the coordinate $x$ normal to the bilayer. $H=0$ (solid line); $H=0.011$ (dashed line); $H=0.016$ (dashed-dotted line). The field is measured in units of $\hbar c/e\xi_S^2$. The other parameters of the bilayer as in Fig.~\ref{TcH1}.}
\label{qual}
\end{figure}

Now we turn to the qualitative discussion of the obtained result. Why $T_c$ is enhanced by the orbital effect of the field? In order to answer this question we first comment on why the FFLO-state is favorable in the dirty S/F bilayer.  The Cooper pairs in the considered FFLO-state $\Delta \propto exp[i\bm k \bm r_{||}]$ have nonzero momentum. For such pairs in the S/F bilayer there are three main depairing factors: (i) effective exchange field; (ii) "leakage" of the superconducting correlations into nonsuperconducting region (F layer); (iii) scattering by the impurities. The nonzero momentum correlations are destroyed by nonmagnetic impurities in contrast to the zero-momentum correlations, which are not sensitive to such scattering according to Anderson theorem. The depairing factor (ii) is not of interest for us now, because it practically does not depend on the momentum of the pair. Which superconducting state is more favorable (homogeneous or FFLO) is determined by the competition of the factors (i) and (iii). From the one hand, the nonzero momentum of the pair can reduce the effective exchange in the bilayer. From the other hand, the larger the pair momentum the stronger the depairing by impurity scattering. The dependence of these depairing factors on $k$ can be extracted from Eq.~(\ref{fS0}) because the anomalous Green's function at zero field can be represented as
\begin{equation}
f_{\sigma,S}^{(0)}\approx \frac{i\pi\Delta}{\varepsilon+\sigma h_{eff}(k)+i\Gamma_0+i\Gamma_1 k^2}
\enspace ,
\label{fS_qual}
\end{equation}
where $\Gamma_0$ stands for the "leakage" of the superconducting correlations into nonsuperconducting region, $\Gamma_1 k^2$ accounts for the impurity depairing and $h_{eff}$ is the effective exchange field. The dependences of $h_{eff}$ and $\Gamma \equiv \Gamma_0+\Gamma_1 k^2$ on $k$ are represented in Fig.~\ref{qual}(a). It is seen that the decrease of the effective exchange in the bilayer can be quite rapid, so the FFLO-state can be realized even in the dirty limit.

Now let us turn on the magnetic field. Then the momentum of the pair should be changed by the generalized momentum $\tilde k=k-(2e/c)A_0-(2e/c)Hx$ (for simplicity we consider the situation of aligned $\bm k$ and $\bm A$, when the effect is maximal). This generalized momentum is a linear function of the coordinate $x$ normal to the bilayer, as represented in Fig.~\ref{qual}(b). It is seen that the average value of $\tilde k$ is enhanced in the F layer: $\langle \tilde k\rangle _F>(k-(2e/c)A_0)_{opt}$ and is reduced in the S layer: $\langle \tilde k\rangle_S<(k-(2e/c)A_0)_{opt}$. It can be shown that the effective exchange field depends only on the generalized pair momentum  in the F layer and the impurity depairing takes place over the whole bilayer, but it is more essential in the superconductor. Therefore, such a modification of the effective pair momentum is favorable for the FFLO-superconductivity for two reasons: (i) it reduces the deparing by the exchange field and (ii) it reduces the depairing by impurity scattering. Consequently, application of the magnetic field in the appropriate direction leads to the enhancement of the critical temperature.

\section{summary} 

In summary, by the example of the bilayer S/F and S/N superconducting systems, which are in the FFLO-state or close to the FFLO instability, we have shown that the orbital effect of the in-plane magnetic field does not necessary suppress superconductivity, but conversely, it can enhance $T_c$. For weak applied fields $T_c$ grows near linearly upon increase of the field, if the system is in the FFLO-state. It is also interesting to investigate if such an effect takes place for other systems (for example, unconventional superconducting films \cite{vorontsov09}), which can be in the modulated superconducting state at zero magnetic field. 

The authors are grateful to A.S. Mel'nikov and S. Mironov for useful discussions.



\begin{thebibliography}{99}
\bibitem{note_zeeman}
We neglect the Zeeman effect of the applied magnetic field. Therefore the mutual orientation of the external field and the exchange field of the ferromagnet can be arbitrary, but the both fields are assumed to be in the plane of the bilayer.
\bibitem{note_kA}
Strictly speaking, the first-order correction is a functional (integral expression) of $\bm k \bm A(x)$, as it is seen from Eq.~(\ref{fS1}).
\end{thebibliography}

\end{document}